# Spatio-temporal propagation of cascading overload failures

Zhao Jichang[1], Li Daqing[2,3]*, Hillel Sanhedrai[4], Reuven Cohen[5], Shlomo Havlin[4]

1. School of Economics and Management, Beihang University, Beijing 100191, China
2. School of Reliability and Systems Engineering, Beihang University, Beijing 100191, China
3. Science and Technology on Reliability and Environmental Engineering Laboratory, Beijing 100191, China
4. Department of Physics, Bar-Ilan University, Ramat Gan 5290002, Israel
5. Department of Mathematics, Bar-Ilan University, Ramat Gan 5290002, Israel

**Different from the direct contact in epidemics spread, overload failures propagate through hidden functional dependencies. Many studies focused on the critical conditions and catastrophic consequences of cascading failures. However, to understand the network vulnerability and mitigate the cascading overload failures, the knowledge of how the failures propagate in time and space is essential but still missing. Here we study the spatio-temporal propagation behavior of cascading overload failures analytically and numerically. The cascading overload failures are found to spread radially from the center of the initial failure with an approximately *constant* velocity. The propagation velocity decreases with increasing tolerance, and can be well predicted by our theoretical framework with one single correction for all the tolerance values. This propagation velocity is found similar in various model networks and real network structures. Our findings may help to predict and mitigate the dynamics of cascading overload failures in realistic systems.**

Resilience of individual components in networks is determined not only by their own intrinsic properties, but also by their functional interactions with other components. For example, a failure of one component in a network may lead to overloads and failures of other components. Starting with a localized failure, such interactions between components can ignite a domino-like cascading failure, which may result in catastrophes such as those observed in many realistic networks [1-5]. The devastating consequences of these cascading failures have stimulated extensive studies [6-9].

Many studies provided deep insight on the conditions and outcome of cascading failures [10-17], which may help to evaluate the system resilience. However, to predict and mitigate the failure spreading in a network, understanding their spatio-temporal propagation properties is essential but still missing. Different from structural cascading failures caused by direct causal dependencies [18-20], overload failures [21-25] usually propagate through invisible paths as a result of cooperative interactions in the system. Actually, a fundamental question has rarely been posed: how overload failures propagate with time in space during the cascading process? Indeed, predicting the spatio-temporal propagation of cascading failures could determine the timing and resource allocation of an effective mitigation strategy in corresponding self-healing technologies.

In this paper, we aim to understand the spatio-temporal propagation of the cascading overload failures in spatially embedded systems. We define two basic quantities to describe the spatio-temporal propagation properties of cascading failures. The first quantity, $r_c(t)$, is the average Euclidean distance of the failures appearing at cascading step $t$ from the center of the initial failure. The second quantity, $F_r(t)$, is the number of node failures that occur at cascading step $t$. While $r_c$ can help system regulators to set a 'firewall' at suitable locations before the failure arrives, $F_r$ can suggest the 'height' of the firewall.

**Model**

In order to study the propagation properties of cascading overload failures, we model the network as a randomly weighted $L \times L$ lattice with periodic boundaries, where $L$ is the linear length of the lattice. The weight of each link is taken from a Gaussian distribution $N(\mu, \sigma^2)$ with mean weight $\mu$ and the disorder is represented by the standard deviation $\sigma$. In this model, the load on a node $i$, $L_i$, represents the number of optimal paths between all pairs of nodes passing through this node [21-22]. A node $i$ will fail when its load $L_i$ is more than $(1+\alpha)$ times its original load, where $\alpha$ represents the system tolerance to overloads [22]. A randomly localized region of the system is initially removed to trigger the cascading overloads. This kind of initial failure is motivated by the fact that natural disasters (like earthquake or floods) or malicious attacks (like WMD) usually occur in specific geographical locations and destroy initially localized regions of the network [25-27]. The initial failure is located in a randomly selected square of $l \times l$ nodes ($l \ll L$), which are removed initially. In addition to being realistic, since failures are usually localized, this configuration can also help us to follow and analyze the spatio-temporal propagation pathway of the cascading failures. This local failure may trigger failures of other nodes if their load value exceeds the tolerance threshold due to the load redistribution across the entire network. Given the periodic boundary conditions, we can position the initial attack region at the center of the lattice.

**Propagation properties of cascading overload failures**

In Fig. 1 we show snapshots of the simulated cascading failures, where the failures spread almost radially from the initial attack region and finally spread over the whole system. As can be seen from Fig. 1, the nodes with spatial location closer to the initial failure begin to fail first and form approximately a 'ring of failures'. The ring begins to grow and expand with time until it reaches the system's boundary.

Considering the ring shape of the cascading overload failures (originating from the initial location), it is reasonable to quantify $r_c(t)$ and $F_r(t)$ at cascading step $t$. Figure 2a shows that the propagation radius of cascading failures is increasing almost linearly with time for different system sizes. This means that the failures spread during the cascade process with an almost constant speed (slope of $r_c(t)$). It can be seen that as the system linear size $L$ increases, the propagation velocity of overload failures increases. The propagation radius of failures is increasing with time and becomes saturated near the boundary. The propagation size of failure $F_r$ (number of new failures at each instant) increases until a certain time step and then decreases (Fig. 2b). The behavior at large $t$ is due to the finite size of the system, when $r_c(t)$ reaches the order of $L$ the amount of damage can only decrease. Note that for different system sizes, the propagation size of failure $F_r$ reaches the maximum at similar instants, which results from the higher velocity in larger system sizes.

To further understand the effect of system size and tolerance $\alpha$ on failure propagation $r_c(t)$, we rescale it in Fig. 2c by the system linear size $L$. We find that the curves of rescaled $r_c$ collapse into a single curve for different system sizes at a given tolerance $\alpha$, suggesting that failures spread in the same relative velocities for different system sizes. When the size of failures, $F_r(t)$, is rescaled by the system size, $L \times L$, the different curves of $F_r(t)/L^2$ also collapse into a single curve. It can be seen that the spreading time needed to reach the maximum $F_r(t)/L^2$ is determined by the tolerance $\alpha$, which implies that large tolerance can postpone the collapse of the system. As seen in Fig. S6 (see SI for more details), there exists a critical value of $\alpha$, $\alpha_c$, above which no spreading occurs. Moreover, in the next section we find the theoretical relation (shown in Fig. S5 in the SI) between the system tolerance and the critical initial failure size, above which the overload failures will spread out. As seen, for a larger $\alpha$, the system can sustain a larger size of initial failure.

In contrast to the dynamics such as epidemics [28] that propagate due to nearest-neighboring interactions such as contact infection, cascading overloads propagate as a result of the global interaction between all the flows contributed by the whole system. Surprisingly, although the interactions are global, the propagation dynamics in the model network and realistic networks (see SI for more examples) are found rather local. Here the local propagation means that there is a finite characteristic distance ($\Delta r_c$) between the successive overloads. This characteristic distance is the value of propagation velocity (see Fig. 5a), which increases non-linearly with decreasing tolerance. The nearest-neighbor propagation of overloads usually assumed in some complex network models only corresponds to the limiting case of this local propagation ($\Delta r_c = 1$).

To further explore these propagation behaviors of cascading failures found in simulations and their relations with tolerance, we develop the following theory.

**Theory**

Cascading overload failures due to an initial local failure are produced by the redistribution of loads in the network. From the observations of simulations on weighted network, we find (see Fig. 1) that failures spread in a ring shape from the center of the initial damage. This inspires us to assume in our theory that the network is embedded in a two dimensional circular plate (see Fig. 3), the initial failure is within a (red) circle of radius *a,* and the main overload due to traffic from a given node *A* around the initial failure is located in the (green) ring adjacent to the initial failure (see the green ring in Fig. 3), whose size will be determined by the theory. The overload is reflected by the increase of the number and lengths of shortest paths passing through this ring. If the overload exceeds the capacity tolerance of a node (($1+\alpha$) times its original load) within the adjacent ring, the node will fail in the next step, causing the cascading failure to propagate forward. Due to the initial failed area, shortest paths from a given node *A* to destinations located in the shadow area *s* (the dotted area in Fig. 3) would be affected and become longer, since they now have to surround the failed area. Specifically, the shortest paths from *A* to nodes in *s* (e.g., *AF*) across the failed area can be separated into two parts in the adjacent ring (e.g., *BC* and *DE*). For the first part, its length within the ring changes from *BC* to *KG*. As for the second part, its overload on the ring can be calculated from symmetry by switching *A* and *F* (source and target). Finally, the integration from $r = a$ to $r = 1$ covers all the overloads added to the adjacent ring (green) due to the initial failure, can be written as

$$\Delta L_r(a,b) = \int_a^b \left[\sqrt{r^2 - a^2}s(a,r) - v(a,b,r)(s(a,r) + \pi a^2)\right]2\pi r dr +$$

$$\int_b^1 \left[\sqrt{b^2 - a^2}s(a,r) - v(a,b,r)(s(a,r) + \pi a^2)\right]2\pi r dr, \quad (1)$$

where the length of *KG* is $\sqrt{b^2 - a^2}$ for *r>b* and $\sqrt{r^2 - a^2}$ for *r≤b*, *s(a, r)* is the

area of *s*, i.e., the number of *A*'s destinations in the shadow and *v(a, b, r)* is the average length of the first part of shortest paths (within the adjacent ring) before the failure. Similarly, we can obtain the initial load of nodes located within the circle centered at *O* with radius *b* as $L_{ini}(b)$ (see Sec. S1.2 in the SI). Then, for a node on the circle centered *O* with radius *b*, the overload produced by the failure is

$$\Delta L = \frac{1}{2\pi b} \frac{\partial}{\partial b} \Delta L_r(a, b), \qquad (2)$$

and its initial load is

$$L_0 = \frac{1}{2\pi b} \frac{\partial}{\partial b} L_{ini}(b). \qquad (3)$$

For each functional node in the network, the critical condition for failure can be written as *α=ΔL/L₀*. Specifically, if *α≥ΔL/L₀*, it survives otherwise it fails. More details for the solution of the theoretical model can be found in the SI.

As can be seen in Fig. 4, our theoretical analysis reproduces well the spatio-temporal propagation features of cascading overloads found in the simulations (Fig. 2). Specifically, as shown in Figs. 4a and 4c, the velocity of the failure propagation is almost constant in most of the time *t*, and decreases with increasing tolerance or decreasing system size. As shown in Figs. 4b and 4d, the number of failures increases with time and then begins to drop after reaching the peak, which is reduced by increasing the tolerance *α*. The instant for the maximal failure size is independent of the system size, but increases with the system tolerance. Moreover, similar to simulation results, both the radius and size of failures in the cascades can also be rescaled with system size (*L* and $L^2$ respectively) as seen in Figs. 4c and 4d, where the different curves for different system sizes collapse into a single curve at a given *α*. Note that both *r_c(t)* and *F_r(t)* demonstrate a small slope in the few initial steps, which is caused by the extremely small initial failure considered in the theory. As shown in the SI (Fig. S3), if the initial size of attack is large, both quantities show a higher slope also in the few initial cascading steps.

To test our theory, we perform a quantitative comparison between the simulation and the theoretical model. As shown in Fig. 5a, the propagation velocity in different model networks and real networks can be well predicted by the theory with the same constant correction for all the values of tolerances. This constant correction, close to 2 π, is a result of the anisotropic propagation of overloads in the simulation, which is assumed isotropic in the theory. These good agreements between theory and simulations support the validity of our proposed theoretical framework for cascading overloads.

The local propagation found in the simulation and theory is due to the mechanism of overload redistribution. In the simulation of the Motter-Lai model in Fig. 5b, the overloads propagate rather locally (within a characteristic distance) after an initial damage due to the redistribution of optimal paths between existing pairs of nodes. The optimal paths that passed through the previous damaged area will now mainly be redistributed close to this area, leading to the local overloads in spite of global interactions. Similar to our findings in the Motter-Lai model, the overloads in the theory (as shown in Fig. 5b) are found here mainly distributed close to the previous damage area, which causes the local propagation of failures. Note the agreement in Fig. 5b between theory and simulations (with the same constant correction of $2\pi$, due to the anisotropic nature of the failure spreading, which is assumed isotropic in the theory). The excellent agreement between simulations and theory of the overload (in Fig. 5b) as a function of distance from the original failure also supports our theoretical approach.

It is worth noting that the analysis of our theoretical framework is not limited by the computational complexity of calculating optimal paths. This is in contrast to simulations based on betweenness (number of optimal paths passing through a node), which require heavy computations and therefore are limited to relatively small systems. Our theoretical framework thus provides an efficient way to explore and understand the cascading behavior of failures for any system size.

**Universality of propagation velocity**

To explore the possibility of universal feature in overload propagation, here we analyze the propagation of cascading overload failures on different network models and realistic networks. Besides the model networks including lattice, circular lattice, planar graph and circular plate, we also study the overload propagation on the realistic road networks [29]. We initiate a local attack in the geographical center of these networks and analyze the spatio-temporal evolution of $r_c$ and $F_r$. As shown in Fig. 5 (see SI for more examples), we find that propagation velocity of overloads is similar in various model and real network structures, which can be well predicted by our theory. The propagation velocity is independent of detailed network structures, which makes our findings more applicable. This universal propagation of overloads can be attributed to the common mechanism of overload redistribution in different networks, independent of structure difference between these networks. Furthermore, both theory and simulations suggest that for a given system size and a given tolerance, the size of initial failure does not influence the spreading velocity (see Fig. S14 in the SI).

**Discussion**

Cascading failures represent the manifestation of nonlinear butterfly effect in infrastructure networks, which can cause catastrophic damages due to a small local disturbance. The Motter-Lai-type overload cascade models are an important class of cascading failure dynamics, characterized by the nonlocal---in contrast to epidemic spreading-type local cascade models---interactions, and have been studied extensively for the last decade. Given the inherent global interactions in the mechanism of overload formation, it is of interest that the overloads spread rather locally from the initial attack region, at velocity that increases non-linearly with decreasing tolerance. Here the local propagation means that there is a finite characteristic distance between the successive overloads, which is the value of propagation velocity. The nearest-neighboring propagation of overloads usually assumed in some complex network models only corresponds to the limiting case of this local propagation (characteristic distance is 1).

For different model and realistic networks, our results suggest the existence of universal propagation features of cascading overloads, which are characterized by a finite linear propagation velocity. This velocity can be predicted by our theory with the same constant correction for all the values of tolerances. This constant correction, close to $2\pi$, is a result of the anisotropic propagation of overloads in the simulation, which is assumed isotropic in the theory. This universal behavior comes from the common global mechanisms of overload redistributions in different networks. When certain extreme heterogeneous networks like embedded scale free networks are considered, a revision of the theoretical framework may be needed. While the focus of this manuscript is on the spatio-temporal propagation of the cascading overload failures in spatially embedded systems, the overloads may spread very fast in general non-spatially-embedded sparse network due to its small diameter.

The present study may help to bridge the longstanding gap between the overload model [16, 22] and the model of dependency links proposed by Buldyrev et al. [13] and Parshani et al. [18], in particular the lattice version of the model [19-20] where dependency links can have a characteristic length $r$. Indeed, as can be seen in Figs. 2 and 4, overload failures propagate in a nearly constant speed, which suggests a characteristic dependency distance, between successive overload failures. Furthermore, this speed or characteristic distance is found to increase with decreasing tolerance. This suggests a possible mapping between systems with overload failures and networks with dependency links, where networks with different characteristic length of dependency links can serve as a suitable model to describe cascading overload failures. This mapping can be useful since overload models usually require heavy computations and are therefore limited to small systems, while dependency models require significantly less computations, and large systems can be easily analyzed.

When a disturbance is detected in networks, the knowledge of spatio-temporal propagation properties of cascading failures is essential for predicting and mitigating the cascading failures. Meanwhile, realistic cascading failures are usually the result of the collective interactions between different processes including overloads and other system operation procedures [30-32]. The universal features of overload propagation found here across different networks may help to better mitigate realistic cascading failure, if combined with the detailed knowledge of other processes including system operations and planning procedures.

**Acknowledgements** We thank the support from the National Natural Science Foundation of China (Grant 61104144) and the National Basic Research Program of China (2012CB725404). D. L. is also supported by Collaborative Innovation Center for industrial Cyber-Physical System. S. H. thanks DTRA, ONR, the LINC and the Multiplex (No. 317532) EU projects, the DFG, and the Israel Science Foundation for support. J. Z. was partially supported by NSFC (Grant Nos. 71501005 and 71531001) and 863 Program (Grant No. SS2014AA012303).

**Author contributions**

All the authors contribute equally.

**Additional information**

Correspondence and requests for materials should be addressed to L. D. (daqingl@buaa.edu.cn).

**Competing financial interests**

The authors declare no competing financial interests.

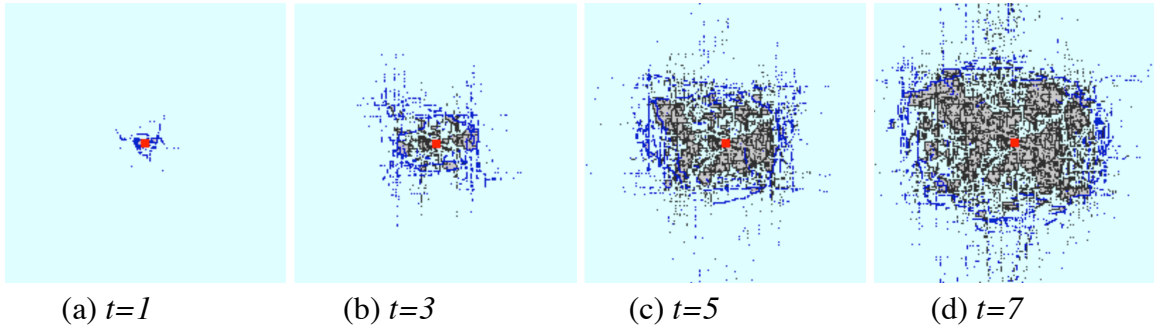

(a) *t=1*  (b) *t=3*  (c) *t=5*  (d) *t=7*

**Figure 1: The propagation of the overload failures in the network.** We demonstrate the step *1, 3, 5* and *7* of the cascading failures on a *200×200* lattice with periodic boundary conditions and a Gaussian distribution of weights. The disorder is $\sigma=0.01$, the initial attack size (in red) is $6 \times 6$, and the tolerance of system is set to $\alpha=0.5$. In each figure, the deep blue dots stands for the overloaded nodes in the current step, while the black ones are the nodes failed in the previous steps. The cyan dots are the functional nodes that did not fail.

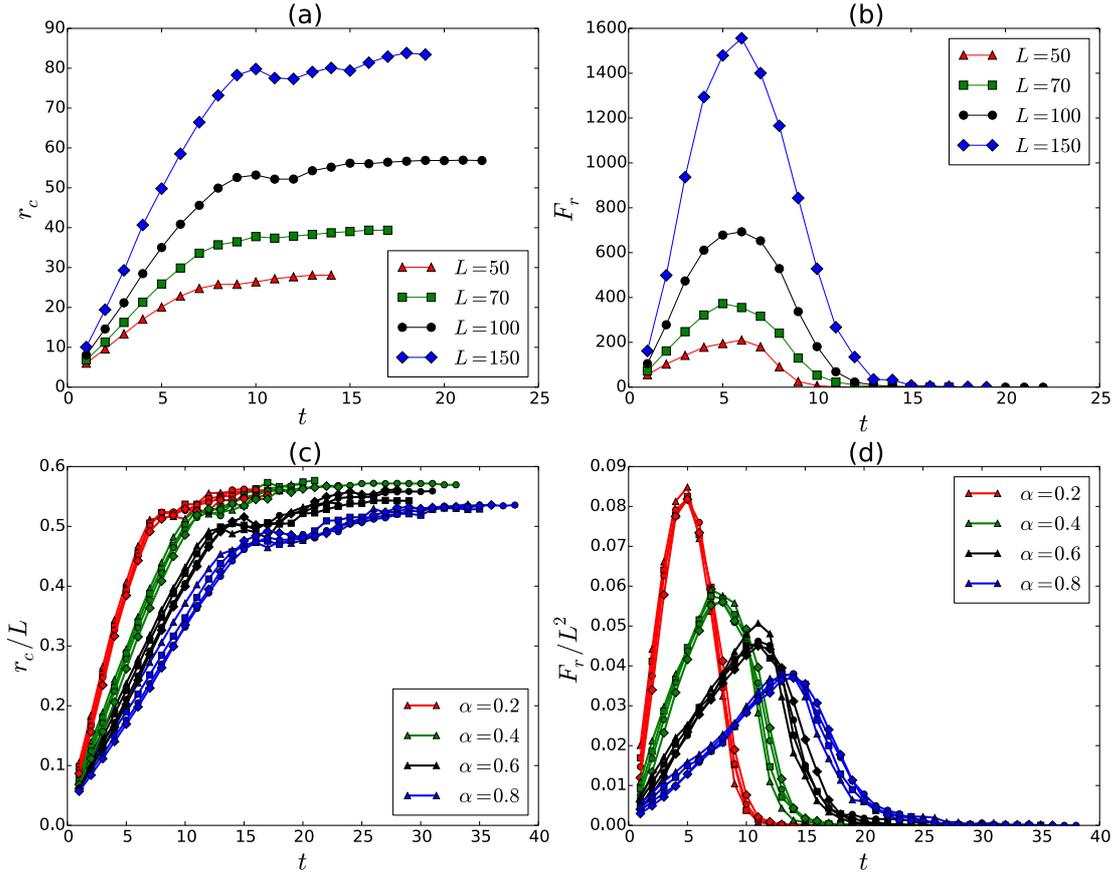

**Figure 2: Spatio-temporal propagation of cascading overload failures in simulations.** (a) and (b) are the spreading radius $r_c(t)$ of failures and amount of failures at each step, $F_r(t)$, as a function of time for $\alpha=0.25$, $\sigma=0.1$ and $l=6$. (c) and (d) are the results of $r_c$ and $F_r$ scaled by the system size, including $L=70$ (triangle), $L=80$ (square), $L=90$ (circle) and $L=100$ (diamond), respectively. Note that simulations are limited by the computational complexity of the most efficient algorithm for calculating node betweenness, which is $O(NM+N^2 \log N)$ for weighted networks, where $N$ is the system size and $M$ is the number of edges ($NM$ is the order of $N^2$ for sparse network).

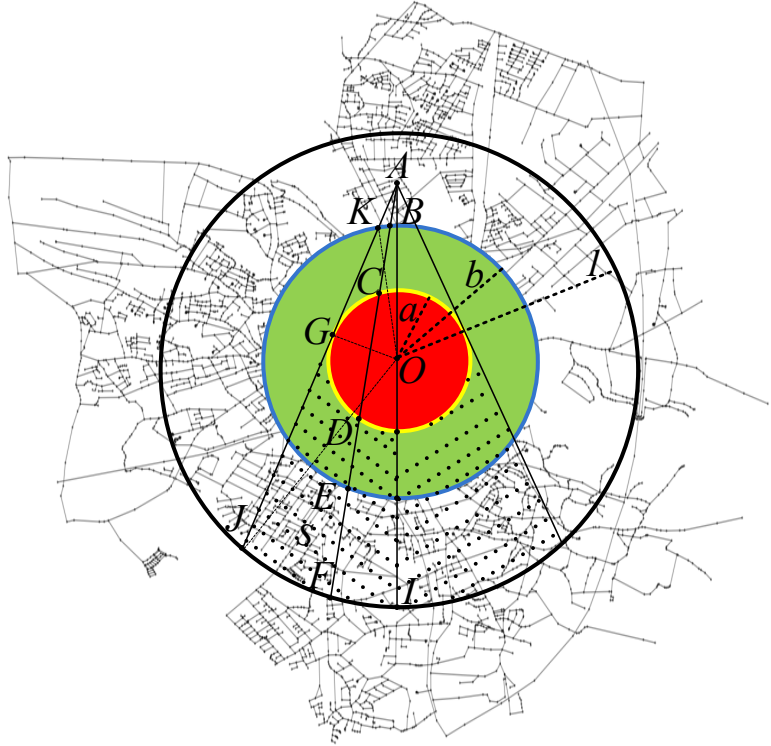

**Figure 3: Theory for overload propagation.** The network is embedded in a *2d* circular plate centered at *O* with a radius of *1* unit and the initial failure is located at the center of the network within a circle of radius $a \ll 1$. The ring centered at *O* and between *a* and *b* (*b>a*) is defined as *the adjacent ring*. *A* is a random node in the network, whose distance to *O* is $r \leq 1$ (here we assume *r>b*, the case of $r \leq b$ can be found in the *SI*). *AF* is the original path starting from *A* to a random node *F* on the system border. Since *r>b*, the intersection points between *AF* and the circle with radius *b* are *B* and *E*. *AF* also intersects the border of the failure area at *C* and *D*. *AJ* is a straight line tangent to the failure area and the tangent point is *G*. We also define another path *AI*, which starts from *A* to a border node *I* by passing through the plate center *O*. Note that a realistic road network is embedded behind the circular plate for demonstration.

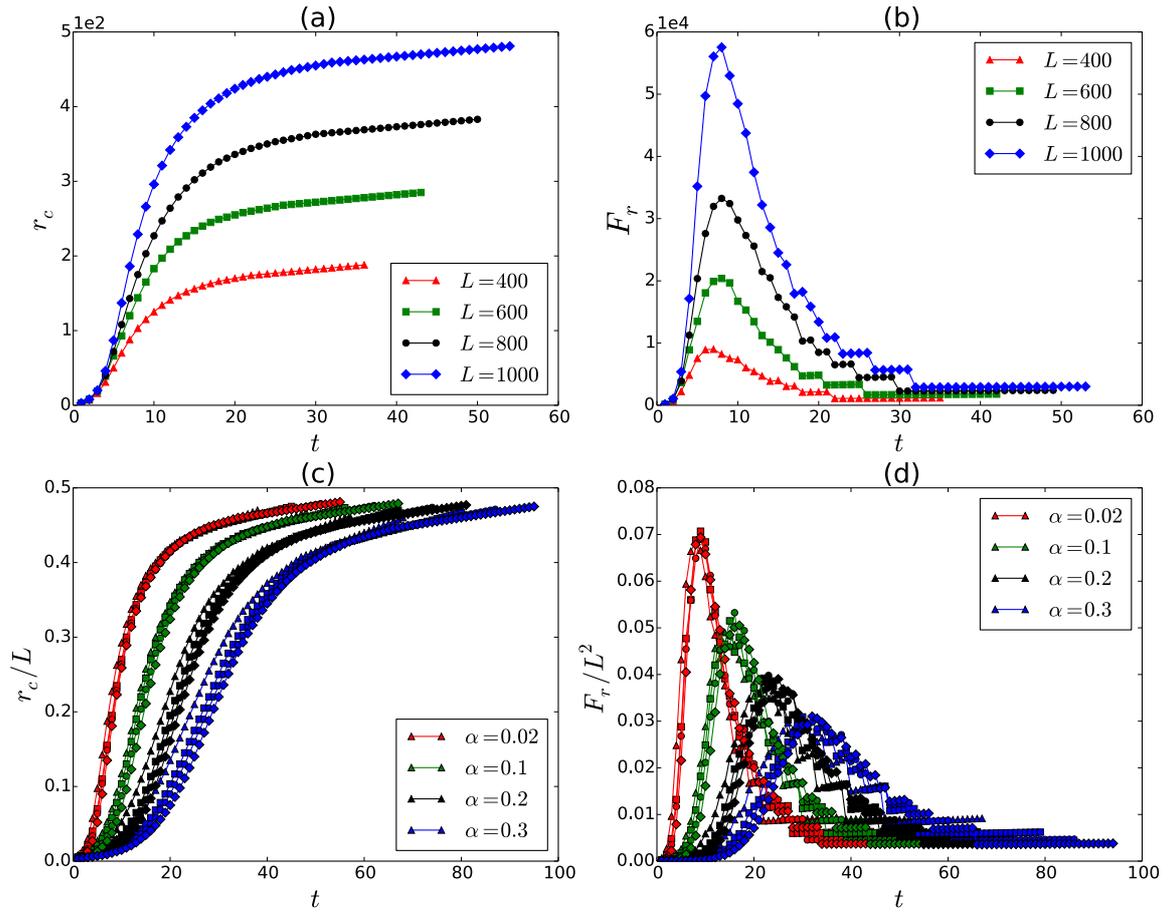

**Figure 4: Theoretical results of overloads propagation.** In (a) and (b), the results from theory with unit distance resolution *1/R* (*L=2R* is the system linear size) are reported for (a) $r_c$ and (b) $F_r$, where *α=0.01*. In (c) and (d), different symbols stand for different resolutions, including *1/200* (triangle), *1/300* (square), *1/400* (circle) and *1/500* (diamond), respectively. Here the absolute radius of the initial damage is set to 3 units.

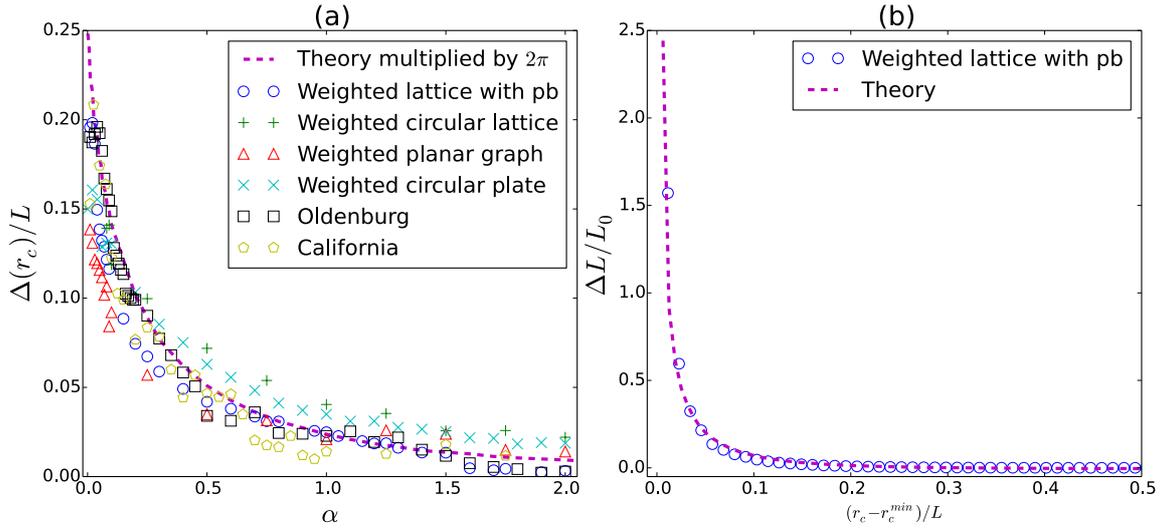

Figure 5: (a) Relative velocity as a function of tolerance in the theory, models as well as in real structures. The relative velocity is calculated in the linear regime of $r_c(t)$, $\Delta(r_c)/L$, which decreases with $\alpha$. The velocity in the theory is multiplied by a constant $2\pi$. We find that the velocity is similar in different model networks (lattice, circular lattice, planar graph and circular plate) and real networks (road network in Oldenburg or California). (b) The average overload as a function of relative distance from the initial attack. The overload in the weighted lattice (circle symbol) and theory (dashed line) after the initial damage is shown. The results are both shifted by the linear size of the initial damage ($r_c^{min}$), which makes two results comparable from the initial attack. We also multiplied the $x$ axis of theory by $2\pi$, which is consistent with the velocity difference shown in (a).